\documentstyle[12pt,aaspp]{article}

\def\pc{\,\hbox{pc}}

\def\msun{{\rm\,M_\odot}}
\def\kpc{{\rm\,kpc}}

\mathchardef\star="313F

\def\Re{\mathop{\it Re}\nolimits}
\def\Im{\mathop{\it Im}\nolimits}

\begin{document}

\title{Weakly Damped Modes in Star Clusters and Galaxies}
\author{Martin D. Weinberg}
\affil{Department of Physics and Astronomy\\University of
Massachusetts, Amherst, \ MA 01003}

\begin{abstract}
A perturber may excite a coherent mode in a star cluster or galaxy.
If the stellar system is stable, it is commonly assumed that such
a mode will be strongly damped and therefore of little practical
consequence other than redistributing momentum and energy deposited by
the perturber.  This paper demonstrates that this assumption is false;
weakly damped modes exist and may persist long enough to have
observable consequences.

To do this, a method for investigating the dispersion relation for
spherical stellar systems and for locating weakly damped modes in
particular is developed and applied to King models of varying
concentration.  This leads to the following remarkable result: King
models exhibit {\it very} weakly damped $m=1$ modes over a wide range
of concentration ($0.67\le c\le1.5$ have been examined).  The
predicted damping time is tens to hundreds of crossing times.  This
mode causes the peak density to shift from and slowly revolve about
the initial center.  The existence of the mode is supported by n-body
simulation.

Higher order modes and possible astronomical consequences are
discussed.  Weakly damped modes, for example, may provide a natural
explanation for observed discrepancies between density and kinematic
centers in galaxies, the location of velocity cusps due to massive
black holes, and $m=1$ disturbances of disks embedded in massive
halos.  Gravitational shocking may excite the $m=1$ mode in globular
clusters, which could modify their subsequent evolution and displace
the positions of exotic remnants.
\end{abstract}


\newpage

\section{Introduction}

A stellar system responds to a disturbance by both phase-mixing, an
incoherent response, and by collective excitation of discrete modes, a
coherent response.  Since galaxy and star cluster models must be
stable to be useful, {\it growing} modes\footnote{For our purposes, a
mode is a well-defined pattern with a possibly complex frequency,
$\omega$.  If $\Im(\omega)>0$, the initial equilibrium is unstable and
the mode initially grows exponentially.  If $\Im(\omega)\le0$, the
equilibrium is stable and the modes are oscillatory or damped.} have
been studied extensively both analytically in the form of stability
criteria and using n-body simulation (see Binney \& Tremaine 1987,
hereafter BT, for a review\nocite{BiTr:87}).  The most well-known of
these is the bar-forming mode which significantly reorganizes its host
galaxy.  In general, {\it damped} modes will also be excited by a
disturbance and transport momentum globally but their amplitude decays
with time.  If the decay time is short, these modes will not be
perceptible. However, if a mode is weakly damped, it may persist for
many dynamical times, even though the system is stable.  For example,
a galaxy ``fly-by'' could excite such a mode which might persist long
after the encounter even if the system is not strictly unstable.
Several authors, in particular Miller and Smith (Miller 1992, Miller
\& Smith 1993\nocite{Mill:92,MiSm:93}), have observed oscillatory
modes in simulations.  Mathur (1990\nocite{Math:90}) formally
demonstrated the existence of oscillating solutions in both
one-dimensional and spherical systems.  In this paper, we explore the
existence and implication of weakly damped modes in spherical stellar
systems.

The existence of damped modes is not in question.  Collective damping
in a homogeneous plasma, {\it Landau damping}, is well-known.  The
simultaneous solution of the Vlasov equation and the Poisson equation
yield a dielectric response function which describes the reaction of
the plasma to an imposed disturbance.  Even in the presence of no
external disturbance, a stochastic space-charge field may give rise to
spontaneous density fluctuations resulting in Landau modes.  The zeros
of the dielectric function are the conditions for collective
solutions.  This use of the response function is sometimes known as
the {\it dispersion relation} (e.g.  Ichimaru 1973\nocite{Ichi:73})
since the it determines the frequency-wavelength relation for
electromagnetic waves in a medium.\footnote{The analogy does not carry
over the stellar dynamical problem but the term {\it dispersion
relation} is still used.} The relationship between the full modal
spectrum and the initial value for the homogeneous plasma has been
discussed by van Kampen (1955\nocite{vanK:55}) and the same physical
principles apply here.

Damped modes are difficult to study in arbitrary stellar systems for
two reasons.  First, the dispersion relation is analytically
intractable in nearly all but the homogeneous cases.  Second, although
one can study damped modes in simulations, one must be able to
discriminate the mode from numerical artifacts.  In this paper, I will
develop a numerical procedure for evaluating the dispersion relation
for spherical stellar systems and locating damped modes (\S2).  Armed
with a prediction for the shape and frequency of a damped mode, we may
initialize and efficiently search for it in n-body simulations.
Spherical models are considered because of the theoretical simplicity,
because they represent a wide variety of astrophysical scenarios and
to relate to the large body of literature on spherical stellar
systems.  In \S3, the technique is applied to King models of varying
concentration.  King models are known to be stable, for example, by
Antonov's criterion (Antonov 1962, see also
BT\nocite{Anto:62,BiTr:87}).  Nonetheless, we will see that these
models all have {\it very} weakly damped $l=m=1$ modes, with damping
times longer than $\sim20$ half-mass crossing times.  This is contrary
to the widely expressed belief that a disturbance in a stable system
will phase mix in several crossing times.  As a check, we realize the
$l=m=1$ mode and follow its evolution in a set of n-body simulations.
The results of the simulation (\S4) show that the expected mode
persists with no obvious decay.  The paper concludes with a discussion
of astronomical applications (\S5).

\section{Method}

For collisionless systems, the dispersion relation describes the
solution to an initial value problem as the result of a Laplace
transform; each component has time-dependence of the form
$\exp(-i\omega t)$, where the frequency $\omega$ may be complex.
Although the dispersion relation for infinite homogeneous stellar
systems has been derived (Ikeuchi et al. 1974,
BT\nocite{IkNT:74,BiTr:87}), most stellar systems are finite and
inhomogeneous. In a stellar sphere in particular, orbits are
quasiperiodic and the repetitive effect of a disturbance at the same
point in an orbit's phase can have a large effect (resonance).
Periodicity may be introduced in a homogeneous medium by carving out a
cube and connecting opposite faces; topologically, this is a 3-torus.
The resulting so-called periodic cube has been used by several authors
(Barnes et al. 1986, Weinberg 1993\nocite{BaGH:86,Wein:93a}) to take
the quasiperiodicity into account without sacrificing the theoretical
simplicity of spatial homogeneity and will be used below as an
example.

Inhomogeneous models of galaxies and clusters are considerably more
difficult to treat analytically because the collisionless Boltzmann
and Poisson equations are separable in different bases.  By Jean's
theorem, the solution to the collisionless Boltzmann equation may be
written conveniently in terms of its integrals of motion.  In the
spherical case, these are energy and angular momentum or radial and
angular actions.  Since the unperturbed Hamiltonian is then
independent of associated angle-like coordinate variables, a linear
perturbation to this equilibrium may be written as a Fourier series in
the coordinate angles.  The Poisson equation is naturally expanded in
a different basis: spherical hamonics.  However, by expanding the
self-consistent gravitational potential in an harmonic series which
satisfies the Poisson equation by construction, the simultaneous
linearized solution to the whole system may be reduced to a matrix
equation.  In particular, let $a_j, b_j$ be two sets of expansion
coefficients.  Then the response $b_j$ to a perturbation $a_j$ is
given by $b_i=M_{ij}(\omega)a_j$ where $M_{ij}$ is the response matrix
and describes the dynamical response of the stellar sphere to that
part of perturbation with time-dependence $\exp(-i\omega t)$.  A
self-consistent response has $a_j=b_j$ (cf.  Appendix B).
This formalism, sometimes called the matrix method, has had varied
applications beginning with Kalnajs (1977\nocite{Kaln:77}) who was
interested in the modes of stellar disks.  Polyachenko and Shukhman
(1981\nocite{PoSh:81}) first adapted the method to the study a
spherical system (see also Fridman and Polyachenko 1984,
Appendix\nocite{FrPo:84b}) and the method was later employed by both
Palmer and Papaloizou (1987\nocite{PaPa:87}) in the study of the
radial orbit instability and by Bertin and Pegoraro
(1989\nocite{BePe:89}) to study the instability of a family of models
proposed by Bertin and Stiavelli (1984\nocite{BeSt:84}).  Weinberg
(1989\nocite{Wein:89}) used the matrix formulation to study the
response of a spherical galaxy to an encounter with a dwarf companion
and Saha (1991\nocite{Saha:91}) and Weinberg (1991\nocite{Wein:91a},
hereafter Paper I) investigated the stability of anisotropic galaxian
models.  Using this approach, the dispersion relation is the condition
that the self-consistent response have a non-trivial solution:
$\det[\delta_{ij}-M_{ij}(\omega)]=0$.  The general dispersion relation
can be calculated for a spherical model with an arbitrary phase-space
distribution function, although only isotropic distributions will be
discussed here.

\begin{figure}[e]
\caption{\label{fig:dispcube} Contours in magnitude of the
analytically-generated dispersion relation for the stellar cube.}
\end{figure}

\begin{figure}[e]
\caption{\label{fig:dispcrat} Same as in Fig. \protect{\ref{fig:dispcube}}
but for the numerically-generated dispersion relation for the stellar
cube.}
\end{figure}

\subsection{Analytic continuation of the dispersion relation}

Just as in the case of the plasma dispersion relation, the derivation
in Paper I explicitly shows that the dispersion relation as written
(see Appendix
B)
is only valid in the upper-half $\omega$ plane.  Causality in the
initial value problem demands that various integrands be analytically
continued to the lower-half plane.  This is readily done in the plasma
case (cf.  Ichimaru 1973\nocite{Ichi:73}) where the natural
coordinates for both the Boltzmann and Poisson equations are
cartesian.  This is problematic in the case of the sphere, whose
dispersion relation is given by equations (\ref{sphmatrix}),
(\ref{potrans}), and (\ref{sphdisp}).  Since equation
(\ref{sphmatrix}) contains the term $1/[\omega-{\bf
n}\cdot{\bf\Omega}({\bf I})]$, explicit analytic continuation requires
analytic continuation of orbital frequencies and actions.  To sidestep
this daunting task, I propose to evaluate the dispersion relation in
the upper-half plane and perform the analytic continuation numerically
by approximating it by a function which may be easily continued.
Given that poles are expected because of the form of equation
(\ref{sphmatrix}), rational functions are a natural choice.

Before applying this to cases of interest, we would like some
assurance that this procedure for analytic continuation will work
reliably.  To do this, I evaluated the dispersion relation for the
periodic cube (eq. \ref{dispcube}), both analytically and numerically.
The simple form of the velocity integrals in equation (\ref{delcube})
allow explicit analytic continuation.  Analytically generated contours
for the fundamental harmonic in the dispersion relation, $|{\cal
D}_{1,0,0}(\omega)|$, are shown in Figure \ref{fig:dispcube}.  The
zeros are at the centers of the concentric contours and the location
of the zeros indicate the type of growth; if the zeros are in the
lower (upper) half-plane, disturbances are damped (growing).  The
model depicted is stable.  The most weakly damped mode has
$\Re(\omega)=0$ and and therefore damps in situ.  The other modes both
damp and propagate with group velocity $\Re(\omega)$.  As the stellar
velocity dispersion is decreased, the solution with $\Re(\omega)=0$
moves toward the real axis from below and will become the Jeans
instability for sufficiently small velocity dispersion.  Modes in
spherical systems, discussed below, have spatial distributions
proportional to spherical harmonics and $\Re(\omega)$ indicates a
pattern speed rather than a group velocity.

I then evaluated ${\cal D}_{1,0,0}(\omega)$ at 20 points for
$-40<\Re(\omega)<40$ for each of $\Im(\omega)=0, 0.5, 1.0$ and found
the corresponding rational function fit (cf. Appendix \ref{app:rat}).
Contours of the approximating rational function are shown in Figure
\ref{fig:dispcrat}.  Comparing Figures
\ref{fig:dispcube} and \ref{fig:dispcrat} we see that the least
damped modes are determined accurately.  I checked the robustness of
the method to the choice of evaluations by varying the number and
placement of points included in the rational function fit.  The three
most weakly damped modes are accurately determined even if any two of
the ``rows'' of constant $\Im(\omega)$ are eliminated.  The most
weakly damped mode remains accurately determined even for 10 points
along the real $\omega$-axis from $-40<\Re(\omega)<40$.  The behavior
of ${\cal D}_{1,0,0}$ near the real axis is dominated by the true
weakly damped modes, allowing zeros to appear at
$\Im(\omega)\approx-28$ in Figure \ref{fig:dispcrat}.  This is an
expected and generic feature; zeros on the lower-half plane will
only be well-approximated by the rational function if they affect the
${\cal D}$ on the upper-half plane.  In general, we can not expect
accurate determination of all features on the lower half-plane from a
finite grid and, in particular, the upper half-plane will place little
constraint on the rational function extrapolation on the lower
half-plane if more weakly damped modes at similar values of
$\Re(\omega)$ exist.

\subsection{Application to the dispersion relation for stellar spheres}

In the next section, we investigate the low-order damped modes for
isotropic King models with dimensionless central potential
$W_0\equiv(E_t-E_o)/\sigma^2=3, 5, 6, 7$ and concentration
$c\equiv\log_{10}(r_t/r_c)=0.67, 1.0, 1.3, 1.5$ respectively where
$r_c$ and $r_t$ are the core and tidal radii (see King
1966 and BT for more
details\nocite{King:66,BiTr:87}).  Linear perturbation theory allows
each harmonic order, $\propto Y_{lm}$, to be considered separately.
However, because the sphere has no symmetry axis, all terms $m$ are
degenerate for a given $l$.  In realizing specific modes below, we
will choose $l=m$.  Units are chosen so that $M=G=1$ and the total
gravitational potential $W=-1/4$.  In most cases, we use a grid in the
upper-half $\omega$-plane with $\Im(\omega)=0.01, 0.05, 0.1$ and 10 to
40 evenly spaced points in $-1<\Re(\omega)<1$.  In some cases the
domain is extended to $-5<\Re(\omega)<5$ to check for ``fast'' pattern
speeds but none were found in any cases explored here.  The
evaluations of ${\cal D}(\omega)$ (cf. eq. \ref{sphdisp}) are fit to a
rational function and zeros in the lower-half plane are noted.  The
robustness of the zeros to the input grid is checked by successively
truncating the grid in $|\Re(\omega)|$, halving the resolution of the
grid and leaving out ``rows'' in $\Im(\omega)$.  In all cases, the
existence of the zeros and general topology of the function did not
change although the position of particular zeros varied by as much as
$10\%$ in a few cases.

The elements of the response matrix $M_{ij}(\omega)$ (cf. eq.
\ref{sphmatrix}) require a two-dimensional integral and a
two-dimensional sum in $l_1$ and $l_2$.  The sum is only two
dimensional since $M_{ij}(\omega)$ is independent of $l_3=m$ given
$l$.  The sum in $l_1$ must be truncated at a chosen value,
$l_{1,max}$ and the sum in $l_2$ has $|l_2|<l$.  Twenty point grids
for the integral evaluation using rational functions were found to be
adequate; large numbers of grid points ($\ga40$) cause significant
truncation error (see App. D).  The rational function technique gives
surprisingly good results with a small number of integrand
evaluations.  Standard lower order integration techniques (e.g.
Simpson's rule) ultimately give better results but only after an order
of magnitude more evaluations.  The value of $l_{1,max}$ was
successively increased to obtain convergence.  The appropriate value
of $l_{1,max}$ depends on $l$; $l_{1,max}$ was chosen to be 6 and 10
for $l=1$ and 2, respectively.  Finally the order of the $n\times n$
matrix $M_{ij}$ was varied to obtain convergence in ${\cal
D}(\omega)$.  The required order increases with $\Re(\omega)$; $n=30$
gave adequate convergence for $|\Re(\omega)|\la1$.  The resulting
error in ${\cal D}(\omega)$ is estimated to be ${\cal O}(10^{-2})$.


\def\sround#1{\hbox to1.6cm{\hfil#1\hfil}}
\def\ssround#1{\hbox to2.5cm{\hfil#1\hfil}}

\begin{table}[t]
\caption{Model parameters and modal frequencies}
\begin{tabular}{ccccccl}
\multicolumn{2}{c}{Model}	& Order
			& \multicolumn{2}{c}{Frequency} & Shift & Example\\[.2ex]
King $W_0$ & $c$	& \sround{$l$}
			& \sround{$\Re(\omega)$}
			& \sround{$\Im(\omega)$}
			& \sround{$r_{shift}/r_c\epsilon$}
			& Cluster name \\
\tableline
3	&0.67	&\sround{1} &\sround{0.010} & \sround{$-0.0016$}
&\sround{0.41}	& NGC 6496	\\
	&	& 2	& 0.10	& $-0.076$	\\
5	&1.0	& 1	& 0.034	& $-0.0011$	& 0.40	& NGC 288 \\
	&	& 2	& 0.0	& $-0.11$	\\
6	&1.26	& 1	& 0.042	& $-0.0065$	& 0.42	& $\omega$ Cen \\
7	&1.5	& 1	& 0.064	& $-0.025$	& 0.54	& M13 \\
	&	& 2	& 0.0	& $-0.20$	\\
\end{tabular}
\end{table}

\section{Properties of damped modes}

This section describes the application of the methods from \S2 in
determining the damped modes of King models with $W_0=3,5,6,7$.  The
complex frequencies of the most weakly damped modes are shown in Table
1.  To summarize, one finds the $l=1$ modes are {\it very} weakly
damped for all $W_0$ studied with the damping proportional to
concentration.  The existence of such modes imply that low-order
excitations of stable galaxies and clusters may persist for many
crossing times and scaled to physical units, perhaps as long as the
age of the galaxy itself.  Very weakly damped $l=2$ modes were found
for the $W_0=3,5$ cases and the same mode is relatively more strongly
damped for $W_0=7$.  No weakly-damped $l=3$ modes were found. A
detailed discussion of these trends and an examination of the modal
shape is given below.

\begin{figure}[e]
\caption{\label{fig:modeloc} Shows the location of the damped $l=1$
modes for $W_0=3,5,6,7$ King models (labeled by $W_0$).  The estimated
precision is indicated by the errorbar at the upper left.}
\end{figure}

\begin{figure}[e]
\caption{\label{fig:mode_w05_1} Contours of the density perturbation
for the $l=m=1$ mode for the $W_0=5$ model on the $x$--$y$ plane.
Only the inner region of the model is shown; the tidal radius is
$r_t=8.72$.  Overdensity (underdensity) is shown as a solid (dashed)
line.}
\end{figure}

\begin{figure}[e]
\caption{\label{fig:comp_w05_1} Effect of the density perturbation
shown in Fig.
\protect{\ref{fig:mode_w05_1}} on the background model.  The peak
perturbation is 30\% of the background.  Note the shift of the central
density peak.  The core radius is at 0.81 in these units.  The dotted
contours show the unperturbed background at the same levels.}
\end{figure}

\begin{figure}[e]
\caption{\label{fig:ek_mode} Absolute value of the phase-space
perturbation for the $l=m=1$ mode for the $W_0=5$ model as a function
of energy $E$ and $\kappa^2=(J/J_{max})^2$.}
\end{figure}

\begin{figure}[e]
\caption{\label{fig:mode_w05_2} Contours of the density perturbation
for the $l=m=2$ mode in the $W_0=5$ model on the $x$--$y$ plane.}
\caption{\label{fig:comp_w05_2} Effect of the density perturbation
shown in Fig. \protect{\ref{fig:mode_w05_2}} on the background model.
The peak perturbation is 30\% of the background.}
\end{figure}

\subsection{Frequencies of lowest order modes} \label{sec:lowest}

For comparison, Figure \ref{fig:modeloc} plots the zeros of the
dispersion relation for $l=1$ damped modes in $W_0=3,5,6$ and $7$
King models.  For example, the $l=1$ harmonic for the $W_0=5$ King
model has a {\it very} weakly-damped mode at $\omega=(\pm0.034,
-0.0011)$.  The $\omega$-plane was searched for
$-1\le\Re(\omega),\Im(\omega)\le1$ and no other modes were
indicated.  Changes in the input grid of ${\cal D}(\omega)$ change the
location of $\Im(\omega)$ for the modes by ${\cal O}(10^{-3})$.  The
errorbar in upper left corner of Figure \ref{fig:modeloc} gives the
estimated precision in $\omega$.  Both the $W_0=3$ and $5$ have nearly
zero damping within the resolution of the technique; the value
$\Im(\omega)=0$ can not be ruled out for these cases.  The analogous
modes for the $W_0=6,7$ models are clearly damped but still only
weakly.

The $l=2$ modes exhibit a similar trend: the rate of damping increases
with concentration.  For the $W_0=5,7$ cases, the pattern speed is
zero but it is nonzero for the lowest concentration model.

Grids for the $l=3$ harmonic were constructed for models $W_0=3, 5, 7$
but no weakly damped modes with $|\Re(\omega)|\la0.3$ and
$\Im(\omega)\ga-0.3$ were found, suggesting that the $l=3$
disturbances are moderately to strongly damped.

\newpage

\subsection{Shape of lowest order modes} \label{sec:shape}

The $l=1$ weakly-damped mode for the $W_0=5$ King model discussed in
\S\ref{sec:lowest} is shown in Figure \ref{fig:mode_w05_1}.  The
procedure for finding the eigenfunction is discussed in Appendix
\ref{app:eigfct}  The peak of the disturbance occurs at $\sim75\%$ of
the core radius of the background model.  The half-mass radius is
$r_{1/2}\approx1.6$.  At larger radii, the density contours remain
concentric.  However, the shape of an $m=1$ mode depends on the
expansion center; therefore one would observe this mode as a coherent
$m=1$ shift at large galactocentric radius by fixing the expansion
center on the density center.

The $l=1$ mode manifests itself as a shift of the central density.
The shift is given by the linear relation
$r_{shift}/r_c\approx0.4\epsilon$, where $\epsilon$ is the ratio of
perturbed density at peak relative to background and
$\epsilon\la0.3$ to have positive density everywhere.  As an example,
the solid contours in Figure
\ref{fig:comp_w05_1} show the combined background and perturbation
with the near maximum amplitude of 30\% at peak.  The dotted contours
show the unperturbed model at the same levels for comparison.  The
shifted center revolves about the original center at angular frequency
$\Re(\omega)$.  The center-of-mass shift for other concentrations is
given in Table 1.  The value $r_{shift}/r_c\epsilon$ is similar for
all concentrations.

The magnitude of the perturbed phase-space distribution averaged over
angles is shown in Figure \ref{fig:ek_mode}.  The energies of
perturbed orbits are small but do not include the most bound particles
and the peak toward large $\kappa^2$ at constant $E$ indicates that
the perturbed orbits tend to be more circular on average.

The $l=m=2$ mode (Fig. \ref{fig:mode_w05_2}) appears as a bar which
peaks at a fraction of the core radius, at roughly the same radial
scale as the $l=1$ mode.  Figure \ref{fig:comp_w05_2} shows the mode
at 30\% amplitude combined with the $W_0=5$ King model background
density.  Since $\Re(\omega)=0$, the pattern will have a fixed
orientation and the bar distortion will decay with time.  The phase
space perturbation is similar in extent to that shown in Figure
\ref{fig:ek_mode} but with a larger contribution of eccentric orbits.

\section{Simulation}

Although the dispersion relation indicates the existence of a weakly
damped $l=1$ mode, can such a mode be excited at significant amplitude?

In partial answer, we may attempt to realize the mode in an n-body
simulation.  Success would also serve as an independent check of the
linearized solution.  Unfortunately, such a simulation is fraught with
serious systematic problems.  First, most simulation schemes do not
explicitly conserve momentum.  Since an $l=1$ perturbation shifts the
density center, numerical artifacts may not be distinguishable from
the mode itself.  To minimize possible artifacts, we use a
direct-summation force calculation which {\it does} conserve
momentum.  Second, minimization of two-body relaxation requires
softening at, roughly, the mean interparticle spacing.  However, the
feature we wish to resolve is only a factor of a few larger than this
scale for $\sim10000$ bodies.  Therefore the softening itself is
likely to change the dispersion relation and this will require that
the dependence on softening be investigated.  Third, even with
softening, relaxation significantly changes the background on a
shorter timescale than the predicted damping, suggesting that
relaxation in the simulation may affect the mode.  Finally, although
these difficulties could be addressed with larger $n$, the $n^2$ FLOP
scaling of the direct summation scheme practically limited the
simulations to $n=10000$.\footnote{Computations were performed on a
Sun Microsystems Sparc 10/41 workstation which required
$3.4\times10^2$ CPU seconds per step.}

\begin{figure}[e]
\caption{\label{fig:virial} The virial quantity $-2T/W$
(top panel) and the run of gravitational potential energy (bottom
panel) shown as a function of time.}
\end{figure}

\begin{figure}[e]
\caption{\label{fig:pos_ang} Position angle as a function of time
for the $l=m=1$ component in the n-body simulation described in \S4.
Three radial wavefunctions (orders $n=8,9,10$) are shown separately
(solid lines) along with the theoretical prediction from \S3 (dashed
line).}
\end{figure}

Because the mode is computed in action-angle variables, the perturbed
n-body phase space is easiest to realize in these variables.  The
details are described in Appendix
\ref{app:nbody} The mode described in
\S3.2
is realized with a peak level of 20\% of background.  The softening is
chosen to be the mean interparticle spacing at the half-mass radius.
Because of both the imposed perturbation and modification of the
interparticle force by softening, the initial conditions are not in
equilibrium.  The system appears to settle at $t=25$
as indicated by a constant value of $-2T/W\sim1$ (cf. Fig.
\ref{fig:virial}).  For comparsion, the radial crossing time at the
half-mass ($\approx1.6$) is roughly 4 time units.  The $-2T/W$
profile looks similar for initial conditions without an initial
perturbation, suggesting that softening causes the poor initial
equilibrium state.

The evolution of the introduced mode may be diagnosed using the same
harmonic decomposition used to evaluate the dispersion relation
(Appendix
B).
The original expansion center is preserved in the center-of-mass frame
since the direct force scheme explicitly conserves momentum.  The
azimuthal phase of the $l=m=1$ distortion for several of the
higher-order radial wavefunctions are shown in Figure
\ref{fig:pos_ang}.  The low-order radial terms are less sensitive to
the mode which has a scale length of $\sim1/10$ the system radius.
The real part of the eigenfrequency is the pattern speed for an $m=1$
distortion.  The dashed line indicates the slope predicted by
$\Re(\omega)$.  The agreement is quite good.  However, the evolution
due to relaxation removes sensitivity to changes in amplitude of the
mode which is predicted to have a longer timescale.

The amplitude of a disturbance may be computed from the total
gravitational potential energy in the disturbance.  From Appendix
B,
it follows that this energy for a particular harmonic is $W_{lm} =
-2\pi G\sum_{j=1}^{n_{max}} |a_j^{lm}|^2$ where $a_j^{lm}$ are the
expansion coefficients.  For the current simulation, the energy per
order $l$, $W_l\equiv\sum_m W_{1m}$, converges with increasing radial
wavenumber $n_{max}$ at fixed time, confirming that the disturbance is
global.  In addition, $W_1$ is roughly constant over the course the
simulation.  More precisely, $W_1$ does appear to grow slightly with
time but this can not be distinguished from the effects of relaxation
which causes the central potential to deepen.  An additional
simulation with a $7\%$ rather than $20\%$ amplitude perturbation
exhibited similar behavior, both in phase and amplitude.

To understand the effects of softening, simulations with softening
length of both half and double the half-mass interparticle spacing
were performed with the same initial conditions.  The same general
trends were seen: the introduced $l=1$ mode persisted with the
expected pattern speed.

However, in all three runs, there were ``bursts'' of the $l=2$ mode,
which might be expected given its long damping time (cf.  Table 1).
The $l=2$ power was larger for smaller softening, possibly suggesting
that either or both relaxation and softening are changing the modal
structure.  In addition, for $t\ga30$, relaxation causes the central
potential to deepen and at the same time, the relative contribution of
the $l=2$ disturbance grows and appears to saturate without damping,
roughly at the level of the initial $l=1$ perturbation.  For the case
with $7\%$ perturbation intially, the $l=1$ mode continues at the
predicted pattern for $t\ga30$ even though the $l=2$ distortion is
evident.  For the $20\%$ case, the growth of the $l=2$ distortion
appears to significantly change the $l=1$ mode.  The importance of
these effects (especially the possible implication for bar growth)
remains to be investigated.

Finally, a set of simulations with no perturbation {\it a priori} were
used to check the significance of the observed disturbance by
comparing the gravitational potential energy in $l=1$ mode.  The
stochastically excited $l=1$ components contained $\la3\%$ of the
energy in the introduced mode described above.  Nevertheless, it is
worth noting that a coherent pattern was seen in these control runs
with the expected pattern speed even though the peak density contrast
was only $1\%$ to $3\%$.

\section{Discussion and Summary}

The numerical evaluation of the dispersion relation for King models
predicts the existence of low-order weakly damped modes.  In
particular, the $l=1$ ($m=1$) mode is {\it very} weakly damped with
small pattern speed over a range of central concentrations.  Similar
results might be expected for other spherical systems.

This mode causes the density center to shift and rotate about the
original center.  Using the linear theory, the shift is given by the
linear relation $r/r_c\approx0.4\epsilon$, where $\epsilon$ is the
ratio of perturbed density at peak relative to the background and must
be $\la30\%$ of background at peak.  This mode has been realized and
observed in a simulation of a $W_0=5$ King model.  A strong excitation
of this mode results in a shift of the density center of roughly
$10\%$ of the core radius as predicted and such levels were maintained
for the duration of the simulation.  However, intrinsic limitations
and features such as softening and relaxation modify the evolution of
and limit temporal sensitivity to weakly damped modes.  This, in part,
explains both the reason they were not discovered by n-body simulation
and the general belief that excitations of stable equilibria vanish in
several crossing times.

Because the low-order damped modes are simple, they should result from
general perturbations.  For example, the differential acceleration due
to a passing galaxy may cause a significant $m=1$ perturbation which
might mix quickly {\it except} for the weakly damped modes.  A
companion on an eccentric orbit may have similar effects.  The induced
distortion due to a companion on circular orbit will be dominated by
the quadrupole ($l=2$) moment; the dipole force is canceled by the
barycentric motion (Weinberg 1989\nocite{Wein:89}).

In the halo of a halo-dominated galaxy, such a mode could result in
the appearance of an $m=1$ distortion or position-shift of the
luminous disk.  Baldwin, Lynden-Bell and Sancisi
(1980\nocite{BLBS:80}) have pointed out that a number of nearby
galaxies have lopsided HI distributions relative to their optical
centers and that a large fraction of spiral galaxies show this
effect.  Damped modes may be partly responsible for the ubiquity of
such offsets, especially in the more moderate cases.  Using the
$W_0=5$ model as an example, the excitation of a $10\%$ perturbation
in a halo with core radius of $15\kpc$ by, say, a companion or
``fly-by'' would result a central density offset of $\sim600\pc$ which
would persist long after the encounter.  In the bulge or spheroid of a
disk galaxy, the mode would shift the density center from the
kinematic center as determined, say, by the large-scale HI rotation
curve.  Using the same example, for a bulge--spheroid with a $1\kpc$
core radius with $\epsilon=0.01,0.1$ the offset would be
$\sim4,40\pc$.  In addition, since the mode is slowly varying, one
might expect a massive central black hole, $M\ga10^6\msun$ to remain
within the potential well of the shifted density center since the
equipartition radius would be $r_c m_\ast/M_{bh}$ for mean stellar
mass, $m_\ast$.

Similarly, a time-dependent external force or {\it gravitational
shock} differentially accelerates globular clusters, possibily
exciting weakly-damped modes.  Clusters on eccentric orbits will
suffer gravitational shocking due to the halo and bulge and all
clusters with mean galactocentric radii within $20\kpc$ will suffer
shocking due to the disk.  In addition, there are relaxation-driven
internal momentum sources such as binary recoil which could in
principle produce an $m=1$ disturbance (Makino \& Sugimoto
1987\nocite{MaSu:87}).  Either way, an offset core may change the
subsequent dynamical evolution by decreasing the thermal contact of
the core and the population of eccentric halo orbits.  This may also
have implications for gravothermal oscillation.  In addition, we may
have to reexamine the intepretation of off-centered compact objects if
any of the $m=1$ excitations keep the density central offset, on
average, from the geometric center.  On the other hand, the physical
picture is further complicated by two-body relaxation whicy may lead
to more rapid damping of these modes.  Also, the largest relative
central density shifts and damping times occur for the lowest
concentrations where we expect rates of relaxation and subsequent
production of exotic objects to be low.  Additional work will be
necessary to understand these competing trends and the extent to which
these modes may be excited in globular clusters.

\acknowledgments
I thank Susan Kleinmann and David van Blerkom for comments on the
manuscript and Dave Chernoff and Ira Wasserman for discussion.  This
work was supported in part by NASA grant NAGW-2224.

\def\astroncite#1#2{#1, #2}

\newpage
\appendix
\setcounter{equation}{0}
\section{Dispersion relation for a stellar cube} \label{app:cudisp}

As discussed in \S2, the dispersion relation is the condition for
simultaneous solution of the collisionless Boltzmann and Poisson
equations in response to a perturbation.  In the case of a homogeneous
cube with periodic boundary conditions, the two equations are
separable in the same coordinate system, cartesian, which leads to an
analytic solution.  Following Barnes et al. (1986)\nocite{BaGH:86},
the dispersion relation may be written:
\begin{equation} \label{dispcube}
	{\cal D}_{\bf n}(p)\equiv 1 + {4\pi GM\over k^2}{\tilde\Delta}_{\bf
n},
\end{equation}
where ${\bf n}$ is a three vector of integers,
\begin{equation} \label{delcube}
	{\tilde\Delta}_{\bf n}\equiv i\int d^3v {{\bf k}\cdot{\partial
	f_o/\partial{\bf v}}\over p+i{\bf k}\cdot{\bf v}},
\end{equation}
$M$ is the total mass of stars in the cube with side length $L$, ${\bf
k}$ is the wave vector defined by ${\bf k}=2\pi{\bf n}/L$, and $f_o$
is the background phase-space distribution function.  The
eigenfunctions are proportional to $\exp(i{\bf k}\cdot{\bf x})$.
Barnes et al.  (1986) derive equation (\ref{dispcube}) by explicitly
introducing a vanishing small growth term.  For an alternative
derivation using the Laplace transform see Weinberg
(1993\nocite{Wein:93a}).

\setcounter{equation}{0}
\section{Dispersion relation for the stellar sphere} \label{app:spdisp}

Any disturbance in the sphere may be represented as a spherical
harmonic expansion with an appropriate set of orthogonal radial
wavefunctions.  We choose a biorthogonal potential-density pair as
described in Paper I.  The pair $(u^{lm}_i,d^{lm}_i)$ is constructed
to satisfy Poisson's equation, $\nabla^2u^{lm}_i = 4\pi Gd^{lm}_i$,
and to form a complete set of functions with the scalar the product
\begin{equation}
	-{1\over4\pi G}\int dr\,r^2 u^{lm\,\ast}_i d^{lm}_j =
	\cases{1& if $i=j$\cr 0& otherwise.\cr}
\end{equation}
The perturbed density and potential then have the following
expansions:
\begin{eqnarray}
	\Phi_1 &=& \sum_{lm}Y_{lm}(\theta,\phi)\sum_j a^{lm}_j(t)
		u^{lm}_j(r),		\label{exp1}	\\
	\rho_1 &=& \sum_{lm}Y_{lm}(\theta,\phi)\sum_j a^{lm}_j(t)
		d^{lm}_j(r).		\label{exp2}
\end{eqnarray}

Paper I shows that the coupled system of collisionless Boltzmann and
Poisson equations may be solved by reducing their simultaneous
solution to a matrix equation using such an expansion above.  The
solution may then be written as the non-trivial solution to the
following linear equation
\begin{equation} \label{sphresp}
	{\tilde a}^{lm}_i =
	M_{ij}(\omega){\tilde a}^{lm}_j,
\end{equation}
where summations over like indices are implied, the `$\tilde{\ }$'
indicates a Laplace-transformed quantity,
\begin{eqnarray} \label{sphmatrix}
	M_{ij}(\omega)
	&=& {(2\pi)^3\over4\pi G}\int\int {dE\,dJ\, J\over\Omega_1(E,J)}
	\sum_{{\bf n}}{2\over2l+1}{\bf n}\cdot{\partial f_0\over\partial{\bf I}}
	{1\over\omega-{\bf n}\cdot{\bf\Omega}}\times \nonumber \\
&&\quad \left|Y_{l,l_2}(\pi/2,0)\right|^2 W^{l_1,\,i\
\ast}_{l,l_2,l_3}({\bf I}) W^{l_1,\,j}_{l,l_2,l_3}({\bf I}),	\\
\noalign{\hbox to\hsize{and\hfill}}
W^{l_1}_{l\,l_2l_3}(I_1,I_2)&=&{1\over\pi}\int^\pi_0 dw_1\cos[l_1w_1-l_2
(\psi-w_2)]u_{j}^{l\,l_3}(r). \label{potrans}
\end{eqnarray}
In the above equations, $I_j$ are the actions, $w_j$ are the angles,
and ${\bf n}=(l_1,l_2,l_3)$ is a vector of integers.  The actions were
chosen by solution of the Hamiltonian-Jacobi equation; $I_1$ is the
radial action, $I_2$ is the total angular momentum and $I_3$ is the
z-projection of the angular momentum.  The quantity $\psi-w_2$, the
difference between the position angle of a star in its orbital plane
and its mean azimuthal angle, depends only on $w_1$, the angle
describing the radial phase.  The frequencies associated with the
conjugate angles ${\bf w}$ are defined by ${\bf\Omega}=\partial
H/\partial{\bf I}$.  $\Omega_1$ is the radial frequency and reduces to
the usual epicyclic frequency for nearly circular orbits.  $\Omega_2$
is the mean azimuthal frequency.  $\Omega_3=0$ and the corresponding
angle $w_3$ is the azimuth of the ascending node (see Paper I or
Tremaine and Weinberg 1984\nocite{TrWe:84}, for details).  It follows
that $M_{ij}^\ast(\omega)=M_{ij}(-\omega^\ast)$ and this property is
exploited in the computations described in \S3 to reduce the number of
numerical evaluations.  Since most researchers quote phase-space
distribution functions in energy and total angular momentum $f(E,J)$,
I have transformed the integration in the definition of ${\bf
M}(\omega)$ in equation (\ref{sphmatrix}) to $E$ and $J$ variables.
The factor $1/[\omega-{\bf n}\cdot{\bf
\Omega}({\bf I})]$ in the integrand of equation (\ref{sphresp}) may
cause discontinuities in the $E$ and $J$ integration.  This is
efficiently addressed using the rational function techniques discussed
in Appendix \ref{app:rat}

Equation (\ref{sphresp}) only has a nontrivial solution if
\begin{equation} \label{sphdisp}
{\cal D}(\omega)\equiv\hbox{det}\{{\bf 1}-{\bf M}(\omega)\}=0.
\end{equation}
As in plasma theory, we refer to the function ${\cal D}(\omega)$ as a
dispersion relation.  In general, $\omega$ will be complex.  Recall
that the coefficients $\tilde{a}_j$ which appear in equation
(\ref{sphresp}) are Laplace transformed and describe the response of
the system to a perturbation with a time-dependence of the form
$\exp(-i\omega t)$.  Because equation (\ref{sphresp}) follows from a
Laplace transform, it is only valid in its present form for $\omega$
in the upper--half complex plane.  The dispersion relation must be
analytically continued to $\Im(\omega)<0$ to find damped modes.  The
response coefficients $\tilde{\bf a}^{lm}$ which solve equation
(\ref{sphresp}) for a particular eigenfrequency $\omega$, describes a
mode of the system.

Because a sphere has no unique symmetry axis, a mode can not depend on
an arbitrary choice of coordinate axes.  Since a rotation causes $m$
components to mix according to the rotation matrices, the mode itself
must depend on $l$ alone.  In particular, note that since the
functions $(u^{lm}_i,d^{lm}_i)$ may be chosen independent of $m$
(e.g.  the spherical Bessel function used by Fridman and Polyachenko
1984, and Weinberg 1989\nocite{FrPo:84b,Wein:89}), equation
(\ref{sphmatrix}) and thus the dispersion relation is independent of
$m$.

\setcounter{equation}{0}
\section{Eigenfunctions} \label{app:eigfct}

The eigenfunction corresponding to the frequency at which ${\cal
D}(\omega)=0$ may be obtained by simultaneous solving the $n-1$
independent equations implied by the $n\times n$ matrix given by
equation (\ref{sphresp}) for the expansion coefficients $a_j$.  The
density and potential perturbations follow directly using equations
(\ref{exp1}) and (\ref{exp2}).  Recall that the eigenvalue is found by
rational function fit to a grid of ${\cal D}(\omega)$.  The matrix
elements at the eigenvalue are determined by rational function fit to
the grid used to find the eigenvalue originally.  The dispersion
relation is not solved exactly by the new interpolated matrix
$M_{ij}(\omega)$ and is iteratively refined to find a new solution to
${\cal D}(\omega)=0$.  Reassuringly the new eigenvalue is found within
5\% in $|\omega|$ of the predicted value in all cases.

\setcounter{equation}{0}
\section{Rational function techniques} \label{app:rat}

Rational functions are used for two tasks in the this paper.  First,
they are suitable estimators for the functional form of the dispersion
relation.  Using a discrete number of evaluations of the dispersion
relation in the upper-half plane, the rational function may be
trivially analytically continued to the lower-half plane and, in
particular, zero locations may be straightforwardly determined.
Second, rational functions are ideally suited to evaluating complex
integrands with poles or near singularities.  Since the integrands of
equation (\ref{sphmatrix}) have ``vanishing''
denominators---denominators of the form $1/[\omega-{\bf n}\cdot{\bf
\Omega}({\bf I})]$---one expects that rational function to approximate
these integrands rather well.

The general procedure is as follows.  Suppose we are given $n$
evaluations of a complex function $f(z_i)$.  The standard techniques
for rational function evaluation may be straightforwardly extended to
the complex domain.  In particular, I adopted reciprocal differences
along the main diagonal (Stoer \& Bulirsch 1980) \nocite{StBu:80} from
which the coefficients of the numerator and denominator polynomials
may be determined directly by recursion.  Let the rational function be
written as
\begin{equation} \label{rat1}
	R(z) = {N(z)\over D(z)},
\end{equation}
where $N$ and $D$ are polynomials and let the order of $N$ be $n$ and
the order of $D$ be $d$.  The reciprocal difference routine gives
either $n=d$ or $n=d+1$.  Using synthetic division,
\begin{equation} \label{rat2}
	R(z) = Q(z) + {{\tilde N}(z)\over D(z)}\equiv Q(z) +
		{\tilde	R}(z)
\end{equation}
where the order of $Q$ and ${\tilde N}$ will be 0 and $n-1$ if $n=d$
or 1 and $n-2$ if $n=d+1$.  Now since ${\tilde R}(\infty)=0$, one can
show that
\begin{equation} \label{ratexp}
	{\tilde	R}(z) = \sum_{i=1}^k S_i(z)
\end{equation}
where the sum is over the $k$ zeros in $D(z)$, $z_i$, and $S_i(z)$ is
the principal part of ${\tilde	R}(z)$ at $z_i$.  Following
Henrici (1974), \nocite{Henr:74} we may write the principal as follows:
\begin{equation} \label{ratterm}
	S_i(z) = \sum_{j=1}^{m_i} a_{i,j}(z-z_i)^{-j}
\end{equation}
where $m_i$ is the multiplicity of the zero at $z_i$.

An efficient method for computing the principle parts has been
outlined by Henrici (1974); \nocite{Henr:74} I will sketch his
arguments here for completeness.  Since the coefficients of $\tilde R$
and $D$ are available, the poles of $D$ may be derived numerically
using deflation and ${\tilde R}(z)$ may be written in the form
\begin{equation}
	{\tilde R}(z) = \sum_{i=1}^k\left({1\over z-z_i}\right)^{m_i} {n(z)\over
d_i(z)}.
\end{equation}
By construction $n(z)/d_i(z)$ is analytic at $z_i$ and may be expanded
in non-negative powers of $h\equiv z-z_i$ and therefore
\begin{equation}
	{\tilde R}(z_i+h) = \sum_{n=0}^\infty c_{i,n} h^{n-m_i}.
\end{equation}
Comparing equation (\theequation) with equations (\ref{ratexp}) and
(\ref{ratterm}), it follows from the uniqueness of Laurent expansion
that $a_{i,j}=c_{i,m_i-j}$ for $j=1,\ldots,m_i$.  The desired values
of $a_{i,j}$ may be then straightforwardly obtained from the explicit
expression of $n(z)$ and $d_i(z)$.

This approach is remarkably useful since any rational function may be
integrated exactly, limited only by truncation errors and the accuracy
of the initial polynomial coefficients.  Therefore, once an expression
in the form of equation (\ref{rat2}) is obtained, the rational
function $R$ maybe be trivially integrated along any segment of the
real line.  In practice, I have found that $m_i$ is rarely larger than
1 and the computational labor is usually much smaller than the
statement of general method might suggest.

One unfortunate feature of the rational function scheme is that does
not converge with increasing grid density due to truncation error.
For example, imagine trying to approximate a lower order rational
function with a very dense grid of points.  The exact representation
will have identical, and therefore canceling, roots in both the
numerator and denominator.  Since these will not cancel precisely in
the numerical representation, a very dense grid may result in a noisy
approximation.  For zero location, incomplete cancellation is easily
recognized.  For integration, one must check the asymptotic behavior.

\setcounter{equation}{0}
\section{N-body realization in action-angle variables} \label{app:nbody}

The most common procedure for realizing an arbitrary but spherical
phase-space distribution is the rejection method for psuedorandom
variables (e.g. Press et al., 1992\nocite{PTVF:92}).  The procedure is
as follows.
Let $\cal R$ be a variate from the unit interval.  If the
phase-space volume is finite for the system, the radius may be computed
from the cumulative mass distribution: $M(r) = {\cal R}M(r_{max})$.
The gravitationally potential $\phi(r)$ then fixes the maximum
velocity.  A trial velocity may then be chosen uniformly from the
sphere of radius $v_{max}=\sqrt{2[\phi(r_{max})-\phi(r)]}$ from which
the energy and angular momentum, $E$ and $J$, may be chosen.  Let
$f_o(E,J)$ be the phase-space distribution function.  The trial point
is accepted if $f_o(E,J)/\sup|f_o(E,J)|>{\cal R}$.  Otherwise the
entire procedure is repeated.

The perturbed phase-space distribution, $f=f_o +\epsilon f_1$, is best
realized in action-angle variables but otherwise the procedure is
analogous.  From Paper I (see also Appendix B), the perturbed mode may be
written
\begin{equation} \label{fpert}
	f_1({\bf I}, {\bf w}) = \sum_{\bf n}
	{\bf n}\cdot{\partial f_0\over\partial{\bf I}}
	{1\over\omega-{\bf n}\cdot{\bf\Omega}}
	e^{i({\bf n}\cdot{\bf w} - \omega t)} \sum_j a_j^{lm}
	W^{l_1,\,j}_{l,l_2,m}({\bf I}).
\end{equation}
One chooses the three trial actions, or equivalently $E, J, J_z$, from
an isotropic distribution, $f_{comp}(E)$, chosen to lie everywhere
above $f$.  The angle variables are chosen uniformly from the interval
$[0,2\pi]$.  The orbit corresponding to the trial phase-space point is
then specified in the background model.  The phase point is accepted
if $f({\bf I}, {\bf w})/f_{comp}(E)>{\cal R}$.  This procedure is
efficient for a careful choice of $f_{comp}$.  The quantity $\epsilon$
must be chosen to be sufficiently small that $f\ge0$ everywhere.

\clearpage

\end{document}